\documentclass{revtex4}
\usepackage{mathrsfs,amssymb,graphicx,epsfig,dcolumn}

\begin{document}
\title{$\delta$ meson effects on neutron stars in the modified quark-meson coupling
model}
\author{Zhong-Ming Niu and Chun-Yuan Gao\\
School of Physics, Peking University, Beijing 100871, China\\E-mail:
gaocy@pku.edu.cn} \maketitle

\begin{center}\begin{minipage}{10cm} \noindent {\bf Abstract}\\\noindent The
properties of neutron stars are investigated by including $\delta$
meson field in the Lagrangian density of modified quark-meson
coupling model. The $\Sigma^-$ population with $\delta$ meson is
larger than that without $\delta$ meson at the beginning, but it
becomes smaller than that without $\delta$ meson as the appearance
of $\Xi^-$. The $\delta$ meson has opposite effects on hadronic
matter with or without hyperons: it softens the EOSes of hadronic
matter with hyperons, while it stiffens the EOSes of pure nucleonic
matter. Furthermore, the leptons and the hyperons have the similar
influence on $\delta$ meson effects. The $\delta$ meson increases
the maximum masses of neutron stars. The influence of
$(\sigma^*,\phi)$ on the $\delta$ meson effects are also
investigated.
\end{minipage}\end{center}
\section{Introduction}
$\delta$ meson is an isovector scalar meson, its contribution is
expected to be neglectable in nuclei with small isospin asymmetry
and in nuclear matter at saturation density. However, for strongly
isospin-asymmetric matter at high densities in neutron stars the
contribution of the $\delta$ field should be considered
\cite{Kubis}. In the last decade, the effects of coupling to the
$\delta$ meson like field on nuclear structure properties of the
drip-line nuclei, on the dynamic situations of heavy ion collisions
and on asymmetric nuclear matter are investigated
\cite{Hofmann,Gaitanos,Liu1,magnetic}. Recently, the density
dependent coupling constants are introduced additionally to
reexamine the $\delta$ meson effects on properties of finite nuclei
and asymmetric nuclear matter in the Quantum Hadron Dynamics (QHD)
model \cite{Haddad,Liu2}. The $\delta$ meson effects are also
investigated in other models, such as a chiral SU(3) model
\cite{chiral}, a relativistic point coupling model \cite{RMFPC},
relativistic transport model \cite{RTM} and so on. But there is no
similar work in the quark-meson coupling (QMC) model yet, so we will
investigate the $\delta$ meson effects by using this model in this
paper.

The quark-meson coupling model was proposed by Guichon in 1988
\cite{Guichon1} where nuclear matter is described as nonoverlapping
MIT bags interacting through the exchange of mesons in the
mean-field approximation. The effective nucleon masses in the QMC
model are obtained self-consistently at the quark level, which is an
important difference from QHD model. The model is refined by
including nucleon Fermi motion and center of mass corrections to the
bag energy by Fleck \cite{Fleck}. Jin and Jennings introduced the
density-dependent bag constant, which is called modified quark-meson
coupling (MQMC) model, to get larger scalar and vector potentials
compatible with experiments \cite{Jin}. Furthermore, the MQMC model
possibly includes the effects of quark-quark correlations associated
with overlapping bags which was missing in the original QMC model,
therefore it is applicable at the densities appropriate to neutron
stars. The $(\sigma^*,\phi)$ meson fields are incorporated to
account for the strong attractive $\Lambda\Lambda$ interaction
observed in hypernuclei which cannot be reproduced by the
$(\sigma,\omega,\rho)$ only in MQMC model \cite{Pal}. The MQMC model
gives a satisfactory description of finite nuclei
\cite{finite_nuclei} and nuclear matter \cite{nuclear_matter}, and
it is widely used in nuclear physics. For example, the temperature
effects of nuclear matter \cite{temperature}, $K$ condensation
\cite{K_condensation1,K_condensation2,Ma}, trapped neutrinos
\cite{trapped_neutrinos}, strong magnetic field
\cite{magnetic_field} and deconfined phenomena \cite{deconfined} in
neutron stars are all investigated in the MQMC model.

In this paper, we extend MQMC model to incorporate $\delta$ meson
field, in which the density-dependent couplings between baryons and
scalar mesons are calculated self-consistently. The model parameters
are determined by the properties of symmetric nuclear matter and
pure neutron matter. Then the influences of leptons, baryons and
$(\sigma^*,\phi)$ mesons on the $\delta$ meson effects are
discussed.

\section{The model}\label{sect2}
The modified quark-meson coupling model is extended to include the
$\delta$ meson field. $\delta$ meson couples only to $u$ and $d$
quarks, because it is built out of nonstrange quarks. $\sigma^*$ and
$\phi$ mesons are also incorporated which couple only to the $s$
quark in a hyperon bag. So there are isoscalar scalar mesons
$\sigma$ and $\sigma^*$, isoscalar vector mesons $\omega$ and
$\phi$, isovector scalar meson $\delta$ and isovector vector meson
$\rho$ in our present model.

In the mean field approximation the Dirac equation for a quark field
of flavor $q\equiv(u,\ d,\ s)$ in the bag for the hadron species
$B\equiv(p,\ n,\ \Lambda,\ \Sigma^+,\ \Sigma^0,\ \Sigma^-,\ \Xi^0,\
\Xi^-)$ is then given by
\begin{equation}
\left[{\rm i}\gamma\cdot\partial-\left(m_
q-g_\sigma^q\sigma-g_{\sigma^*}^q\sigma^*
-g_\delta^qI_{3q}\delta_3\right) -\gamma^0\left(g_{\omega}^q\omega_0
+g_{\phi}^q\phi_0+g_{\rho}^qI_{3q}\rho_{03}
\right)\right]\psi_{qB}(\vec{r},t)=0.
\end{equation}
Here $I_{3q}$ is the isospin projection of quark $q$; $g_\delta^q$
is the coupling constant between quark $q$ and $\delta$ meson,
$\delta_3$ denotes expectation value of the isospin 3rd-component of
$\delta$ meson field, and the other symbols are the same as in
\cite{Ma}. The normalized ground state is solved as
\begin{equation}\psi_{qB}=\mathcal{N}_{qB}\exp\left(\frac{-{\rm i}\epsilon_{qB}t}{R_B}\right)
\left(
\begin{array}{c}
j_0\left(\displaystyle\frac{x_{qB}r}{R_B}\right) \\
\displaystyle{\rm i}\beta_{qB}\vec{\sigma}\cdot\hat{r}j_1
\left(\displaystyle\frac{x_{qB}r}{R_B}\right)
\end{array}
\right)\frac{\chi_{qB}}{\sqrt{4\pi}}
\end{equation}
where
\begin{eqnarray}
\epsilon_{qB}&=&\Omega_{qB}{\pm}R_B\left(g_{\omega}^q\omega_0
+g_{\rho}^qI_{3q}\rho_{03}+g_{\phi}^q\phi_0\right),\label{3}\\
\beta_{qB}&=&\sqrt{\frac{\Omega_{qB}-R_Bm_q^*}{\Omega_{qB}+R_Bm_q^*}},\\
\Omega_{qB}&=&\sqrt{x_{qB}^{2}+\left(R_Bm_q^*\right)^{2}},
\end{eqnarray}
with $R_B$ is the bag radius of baryon $B$ and $x_{qB}$ is the
dimensionless quark momentum which can be determined by the linear
boundary condition
\begin{equation}j_0(x_{qB})=\beta_{qB}j_{1}(x_{qB})\end{equation} The effective
quark mass is
\begin{equation}m_q^*=m_q-g_\sigma^q\sigma
-g_{\sigma^*}^q\sigma^*-g_\delta^qI_{3q}\delta_3\end{equation} The
energy of a MIT bag for baryon $B$ is then given by
\begin{equation}
E_B^{\rm
bag}=\frac{\displaystyle\sum_qn_{qB}\Omega_{qB}-z_B}{R_B}+\frac{4}{3}\pi
R_B^{3}B_B\left(\sigma,\sigma^*,\delta_3\right)
\end{equation}
where $n_{qB}$ is the number of constituent quark $q$ in baryon $B$,
$z_B$ is the zero-point motion parameter and $B_B$ is the medium
dependent bag parameter. The ansatz for the coupling of bag
parameter to the scalar fields $\sigma$, $\sigma^*$ \cite{zj} is
extended to $\delta_3$
\begin{equation}
B_B\left(\sigma,\sigma^*,\delta_3\right)=B_0\exp\left\{-\frac{4}{M_B}
\left[n_{sB}g_{\sigma^*}^{\rm
bag}\sigma^*+\sum_{q=u,d}n_{qB}\left(g_\sigma^{\rm
bag}\sigma+g_\delta^{\rm bag}I_{3q}\delta_3\right)\right]\right\}
\end{equation}
where $B_0$ is the bag constant in free space, $M_B$ is the bare
mass of the baryon $B$, and $g_\sigma^{\rm bag}$, $g_{\sigma^*}^{\rm
bag}$ and $g_\delta^{\rm bag}$ are real parameters.

After the corrections of spurious center of mass motion, the
effective baryon mass is given by
\begin{equation}\label{eq:mstar}
M_B^*=\sqrt{\left(E_B^{\rm bag}\right)^{2}-\left<p_{\rm
c.m.}^{2}\right>_B}
\end{equation}
where
\begin{equation}
\left<p_{\rm
c.m.}^{2}\right>_B=\frac1{R_B^{2}}\sum_qn_{qB}x_{qB}^{2}
\end{equation}
The bag radius $R_B$ could be obtained through the minimization of
the baryon mass with respect to the bag radius
\begin{equation}\label{minimization}
\frac{\partial M_B^*}{\partial R_B}=0
\end{equation}

Consider an many-particle system consisting of the full baryon octet
which interact via $\sigma, \sigma^*, \omega, \phi, \delta, \rho$
meson fields. The Lagrangian density
is\begin{eqnarray}\mathcal{L}&=&\sum_{B}\bar{\Psi}_{B}\left[{\rm
i}\gamma_\mu\partial^\mu-M_{B}^{*}\left(\sigma,\sigma^*,\delta_3\right)
-\left(g_{\omega B}\omega_\mu\gamma^\mu+g_{\rho
B}\frac{\vec{\tau}_B}{2}\cdot{\vec{\rho}}_\mu\gamma^\mu +g_{\phi
B}\phi_\mu\gamma^\mu\right)\right]\Psi_{B}
+\frac12\left(\partial_\mu\sigma\partial^\mu\sigma\right.\nonumber\\
&&\left.+\partial_\mu\vec{\delta}\cdot\partial^\mu\vec{\delta}
+\partial_\mu\sigma^*\partial^\mu\sigma^*\right)
-\frac{1}{2}\left(m_{\sigma}^{2}\sigma^{2}
+m_\delta^2\vec{\delta}\cdot\vec{\delta}+m_{\sigma^{*}}^{2}\sigma^{*2}
-m_{\omega}^{2}\omega_\mu\omega^\mu
-m_{\rho}^{2}{\vec{\rho}}_\mu\cdot{\vec{\rho}}^\mu
-m_{\phi}^{2}\phi_\mu\phi^\mu\right)\nonumber\\
&&-\frac14\left(W_{\mu\nu}W^{\mu\nu}
+\vec{G}_{\mu\nu}\cdot\vec{G}^{\mu\nu}+F_{\mu\nu}F^{\mu\nu}\right)+\sum_{l}\bar{\Psi}_{l}\left({\rm
i}\gamma_{\mu}\partial^{\mu}-m_{l}\right)\Psi_{l}\label{eq:L}
\end{eqnarray}
where $l\equiv(e,\mu)$. Then from Eq. (\ref{eq:mstar}) and
(\ref{eq:L}), we can derive the equations of the motion for the
meson fields in uniform static matter:
\begin{eqnarray}
\label{fields}m_\sigma^{2}\sigma&=&\frac1{\pi^{2}}\sum_Bg_{\sigma
B}C_B(\sigma)\int_0^{k_B}\frac{M_B^*}
{\left[k^{2}+M_B^{*2}\right]^{1/2}}k^{2}{\rm d}k,\\
\label{fieldss}m_{\sigma^*}^{2}\sigma^*&=&
\frac1{\pi^{2}}\sum_Bg_{\sigma^* B}C_B\left(\sigma^*\right)
\int_0^{k_B}\frac{M_B^*}
{\left[k^{2}+M_B^{*2}\right]^{1/2}}k^{2}{\rm d}k,\\
\label{fieldd}m_\delta^{2}\delta_3&=&\frac1{\pi^{2}}\sum_Bg_{\delta
B} C_B(\delta_3)\int_0^{k_B}\frac{M_B^*}
{\left[k^{2}+M_B^{*2}\right]^{1/2}}k^{2}{\rm d}k,\\
\label{fieldo}m_{\omega}^{2}\omega_0&=&
\frac1{3\pi^{2}}\sum_Bg_{\omega B}k_B^{3},\\
\label{fieldp}m_{\phi}^{2}\phi_0&=&\frac1{3\pi^{2}}\sum_Bg_{\phi B}k_B^{3},\\
\label{fieldr}m_{\rho}^{2}\rho_{03}&=&\frac1{3\pi^{2}}\sum_Bg_{\rho
B}I_{3B}k_B^{3}.
\end{eqnarray}
Here $k_B$ is the Fermi momentum of the baryon species $B$. The
factors $C_B(\sigma), C_B(\sigma^*), C_B(\delta_3)$ are:
$$g_{\phi B}C_B(\phi)=-\frac{\partial
M_B^*}{\partial\phi},\hskip1cm\phi=\sigma,\ \sigma^*,\
\delta_3$$
\begin{eqnarray} -\frac{\partial M_B^*}{\partial\sigma}
&=&\frac{E_B^{\rm
bag}}{M_B^*}\sum_{q=u,d}n_{qB}\left\{g_\sigma^q\left[S_{qB}\left(1-\frac{\Omega_{qB}}{E_B^{\rm
bag}R_B}\right)+\frac{m_q^*}{E_B^{\rm bag}}\right]+\frac{16\pi
g_\sigma^{\rm
bag}B_BR_B^{3}}{3M_B}\right\}\label{g1}\\
-\frac{\partial M_B^*}{\partial\sigma^*} &=&\frac{E_B^{\rm
bag}}{M_B^*}n_{sB}\left\{g_{\sigma^*}^q
\left[S_{sB}\left(1-\frac{\Omega_{sB}}{E_B^{\rm
bag}R_B}\right)+\frac{m_s^*}{E_B^{\rm bag}}\right]+\frac{16\pi
g_{\sigma^*}^{\rm
bag}B_BR_B^{3}}{3M_B}\right\}\label{g2}\\
-\frac{\partial M_B^*}{\partial\delta_3} &=&\frac{E_B^{\rm
bag}}{M_B^*}\sum_{q=u,d}n_{qB}I_{3q}\left\{g_\delta^q
\left[S_{qB}\left(1-\frac{\Omega_{qB}}{E_B^{\rm
bag}R_B}\right)+\frac{m_q^*}{E_B^{\rm bag}}\right]+\frac{16\pi
g_\delta^{\rm bag}B_BR_B^{3}}{3M_B}\right\}\label{g3}
\end{eqnarray}
The scalar density of quark $q$ in the bag $B$ are
\begin{equation}\label{S}
S_{qB}=\frac{\Omega_{qB}/2+R_Bm_q^*(\Omega_{qB}-1)}{\Omega_{qB}(\Omega_{qB}-1)
+R_Bm_q^*/2},\quad q\equiv(u,d,s),
\end{equation}

At last, there are two conditions left:
\begin{eqnarray}
\label{neutrality}&&\textrm{charge
neutrality:}\,\sum_Bq_Bk_B^{3}=\sum_lk_l^{3};\\
\label{betaequ}&&\textrm{$\beta$
equilibrium:}\,\mu_B=\mu_n-q_B\mu_e,\quad \mu_{\mu}=\mu_e.
\end{eqnarray}
where $q_B$ and $\mu_B$ correspond to the electric charge and
chemical potential of baryon $B$, respectively. The energy
eigenvalue of Dirac equation for baryon $B$ and lepton $l$ are:
\begin{eqnarray}
\epsilon_B&=&\sqrt{k_B^{2}+M_B^{*2}}+g_{\omega
B}\omega_0+g_{\phi B}\phi_0+g_{\rho B}I_{3B}\rho_{03},\label{eB}\\
\epsilon_l&=&\sqrt{k_l^{2}+m_l^{2}}\label{el}
\end{eqnarray}
Then the Fermi momentum can be obtained from the equations
\begin{eqnarray}\epsilon_B(k_B)&=&\mu_B\label{BaryonFermi}\\
\epsilon_l(k_l)&=&\mu_l\label{LeptonFermi}\end{eqnarray}

After the meson fields ($\sigma, \sigma^*, \omega, \phi, \delta_3,
\rho_{03}$), Fermi momenta ($k_B, k_l$) and effective masses $M_B^*$
are obtained by solving the Eqs. (\ref{fields})--(\ref{fieldr}),
(\ref{BaryonFermi})--(\ref{LeptonFermi}) and (\ref{minimization})
self-consistently at a given baryon number density
\begin{equation}\rho=\frac1{3\pi^{2}}\sum_Bb_Bk_B^{3}\label{totaldensity}\end{equation}
where $b_B$ is the baryon number of baryon $B$, we can obtain the
total energy density and pressure:
\begin{eqnarray}
\varepsilon&=&\frac{1}{2}\left(m_\sigma^{2}\sigma^{2}
+m_{\sigma^*}^{2}\sigma^{*2}+m_{\omega}^{2}
\omega_0^{2}+m_{\phi}^{2}\phi_0^{2}
+m_\delta^{2}\delta_3^2+m_{\rho}^{2}\rho_{03}^{2}\right)\nonumber\\
&&+\frac1{\pi^{2}}\sum_B\int_0^{k_B}\left[k^{2}
+M_B^{*2}\right]^{1/2}k^{2}{\rm
d}k+\frac{1}{\pi^{2}}\sum_l\int_0^{k_l}\left[k^{2}+m_l^{*2}\right]^{1/2}k^{2}{\rm
d}k
\label{energydensity}\\
P&=&\frac{1}{2}\left(m_{\omega}^{2}\omega_0^{2}
+m_{\phi}^{2}\phi_0^{2}+m_{\rho}^{2}\rho_{03}^{2}
-m_\sigma^{2}\sigma^{2}
-m_{\sigma^*}^{2}\sigma^{*2}-m_\delta^{2}\delta_3^{2}\right)\nonumber\\
&&+\frac1{3\pi^{2}}\sum_B \int_0^{k_B}\frac{k^{4}{\rm
d}k}{\left[k^{2}+M_B^{*2}\right]^{1/2}}+\frac{1}{3\pi^{2}}\sum_l\int_0^{k_l}\frac{k^{4}{\rm
d}k}{\left[k^{2}+m_l^{*2}\right]^{1/2}}\label{pressure}
\end{eqnarray}
\section{Parameters and calculation details}\label{sect3}
Take the current quark mass to be $m_u=m_d=0$ and $m_s=150$ MeV.
Small current quark mass for the non-strange flavors or other values
for the strange flavor lead only to small numerical refinements
\cite{K_condensation2}. The meson masses are $m_\sigma=550$ MeV,
$m_\sigma^*=980$ MeV, $m_{\rho}=775$ MeV, $m_\delta=985$ MeV,
$m_{\omega}=783$ MeV, $m_{\phi}=1020$ MeV, respectively.

Assume $\sigma, \omega, \rho, \delta$ mesons couple only to the $u$,
$d$ quarks and $\sigma^*, \phi$ mesons couple only to the $s$ quark,
we have
\begin{equation}
g_\sigma^s=g_{\omega}^s=g_\delta^s=g_{\rho}^s
=g_{\sigma^*}^u=g_{\sigma^*}^d=g_{\phi}^u=g_{\phi}^d=0
\end{equation}
By assuming the SU(6) symmetry of the simple quark model\cite{Pal}
\begin{equation}\begin{array}{llllllll}&g_\sigma^d=g_\sigma^u,&
g_{\sigma^*}^s=\sqrt2g_{\sigma}^u, &g_\delta^d=g_\delta^u;
&g_\omega^d=g_\omega^u,& g_{\phi}^{s}=\sqrt2g_{\omega}^u,
&g_\rho^d=g_\rho^u& \end{array}\end{equation} we can get the
relations
\begin{eqnarray}
&&\frac{1}{3}g_{\omega
N}=\frac{1}{2}g_{\omega\Lambda}=\frac{1}{2}g_{\omega\Sigma}=g_{\omega\Xi}=g_\omega
^u\\
&&g_{\phi\Lambda}=g_{\phi\Sigma}=\frac{1}{2}g_{\phi\Xi}=\sqrt2g_{\omega}^u,\hskip5mm
g_{\phi N}=0\\
&&g_{\rho N}=g_{\rho\Lambda}=g_{\rho\Sigma}=g_{\rho\Xi}=g_\rho^u
\end{eqnarray}
To reduce parameters we set\begin{equation}\frac{g_{\delta}^{\rm
bag}}{g_{\delta}^u}=\frac{g_{\sigma^*}^{\rm
bag}}{g_{\sigma^*}^s}=\frac{g_{\sigma}^{\rm
bag}}{g_{\sigma}^u}\end{equation}

The free nucleon zero-point motion parameter $z_{N0}$ and the free
bag constant $B_0$ are fixed to reproduce the free mass of nucleon
$m_{N}=939$ MeV with the minimization condition (\ref{minimization})
at a free bag radius $R_{N0}=0.6$fm. Then the free zero-point motion
parameters $z_{B0}$ and free radii $R_{B0}$ of other baryons are
obtained by reproducing the free baryon mass $M_B$ with the
minimization condition (\ref{minimization}). They are all listed in
the Table \ref{tb1}.
\begin{table}
\begin{center}
\caption{The zero-point motion parameters $z_{B0}$ and bag radii
$R_{B0}$ in free space are obtained to reproduce the free space mass
spectrum after the parameters $B_0^{1/4}=188.102$ MeV and
$z_{N0}=2.030$ have been fixed by the properties of
nucleon.}\begin{tabular} {c|ccc}\hline\hline
&$M_B$(MeV)  &$z_{B0}$  &$R_{B0}$(fm) \\
\hline
$\Lambda$ &1115.68  &1.815  &0.643\\
$\Sigma^+$ &1189.37  &1.638 &0.669\\
$\Sigma^0$ &1192.64  &1.630 &0.670\\
$\Sigma^-$ &1197.45  &1.612 &0.672\\
$\Xi^0$ &1314.83 &1.501 &0.689\\
$\Xi^-$ &1321.31 &1.483 &0.689\\
\hline \hline
\end{tabular}\label{tb1}
\end{center}
\end{table}

\begin{figure}
\begin{center}
\includegraphics[width=6.5cm]{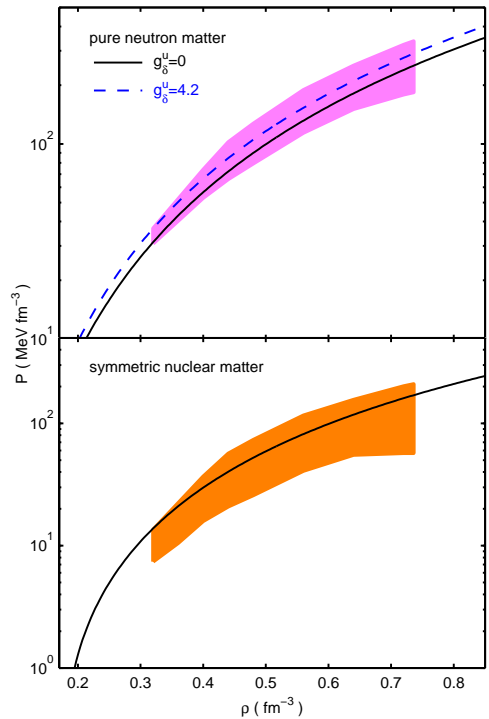}
\caption{(Color online) The EOSes obtained in MQMC model for pure
neutron matter and symmetric nuclear matter. The upper magenta
hatched area and the lower orange hatched area correspond to the
pressure regions for neutron matter after inclusion of the pressure
from asymmetry term with strong density dependence and for symmetric
nuclear matter consistent with the experimental flow data,
respectively \cite{g_delta}.} \label{gdelta}
\end{center}
\end{figure}

Four independent coupling constants $g_\sigma^u, g_{\omega}^u,
g_{\rho}^u$ and $g_\sigma^{\rm bag}$ can be adjusted by reproducing
the symmetric nuclear matter binding energy $B/A$=16 MeV, symmetry
energy $a_{\rm sym}$=32.5 MeV and compressibility $K=289$ MeV at
saturation density $\rho_0=0.17$ ${\rm fm}^{-3}$, as listed in Table
\ref{tb2}. The $\delta$ meson-quark coupling constant is constrained
in an range of $0\le g_\delta^u\le4.2$ so that the pure neutron
matter EOS is consistent with the experimental flow data in
heavy-ion collision \cite{g_delta}, which is shown in the upper
panel of Figure. \ref{gdelta}, the EOS for symmetric nuclear matter
is also shown in the lower panel and we can see that it is also
consistent with the experimental flow data in heavy-ion collision.
\begin{table}
\begin{center}
\caption{Four independent coupling constants are fixed to reproduce
the symmetric nuclear matter binding energy $B/A=16$ MeV, symmetry
energy $a_{\rm sym}=32.5$ MeV and compressibility $K=289$ MeV at
saturation density $\rho_0$=0.17 fm$^{-3}$. $\delta$ meson coupling
constant is set 0 and 4.2.}\begin{tabular}{c|cccc}\hline\hline \ \
$g_\delta^u$\ \ &\ \ $g_\sigma^u$\ \  &\ \ $g_{\omega}^u$\ \ &\ \
$g_{\rho}^u$\ \ &\ \ $g_\sigma^{\rm
bag}$\ \ \\
\hline
0 &0.980  &2.705  &7.948 &2.278\\
4.2 &0.980  &2.705 &10.217 &2.278\\
\hline \hline
\end{tabular}\label{tb2}
\end{center}
\end{table}

The equilibrium properties of neutron stars are obtained by solving
Tolman-Oppenheimer-Volkoff (TOV)\cite{TOV} equations\begin{eqnarray}
\frac{{\rm d}P(r)}{{\rm
d}r}&=&-\frac{G\left[\varepsilon(r)+P(r)\right]\left[M(r)+4\pi
r^{3}P(r)\right]}{r^{2}\left[1-2GM(r)/r\right]}\\ \frac{{\rm
d}M(r)}{{\rm d}r}&=&4\pi r^{2}\varepsilon(r).
\end{eqnarray}
The Baym-Pethick-Sutherland model \cite{BPS} is used to describe the
EOS at subnuclear densities.
\section{Results and discussions}\label{sect4}
Four cases in Table \ref{tb3} are investigated: (1) pure neutron
matter denoted by $nn$; (2) $\beta$-equilibrium nucleonic matter
denoted by $np$; (3) $\beta$-equilibrium hadronic matter composed of
baryon octet without $(\sigma^*, \phi)$ meson fields, denoted by
$npH$; (4) The same as in Case (3) with two additional meson fields
$(\sigma^*, \phi)$, denoted by $npH^*$.
\begin{table}
\begin{center}
\caption{The cases we study in the paper. $H$ represents hyperons
$(\Lambda, \Sigma^+, \Sigma^0, \Sigma^-, \Xi^0,
\Xi^-)$.}\begin{tabular} {c|c|c|c|c}\hline\hline
notation &$nn$ &$np$ &$npH$ &$npH^*$\\
\hline
baryons &$n$ &$n, p$  &$n, p, H$  &$n, p, H$\\
\hline
leptons &  &$e$  &$e$, $\mu$ &$e, \mu$\\
\hline mesons &$\sigma, \omega, \rho(, \delta)$  &$\sigma, \omega,
\rho(, \delta)$  &$\sigma, \omega,
\rho(, \delta)$  &$\sigma, \omega, \rho(, \delta), \sigma^*, \phi$\\
\hline \hline
\end{tabular}\label{tb3}
\end{center}
\end{table}

\begin{figure}
\begin{center}
\includegraphics[width=7cm]{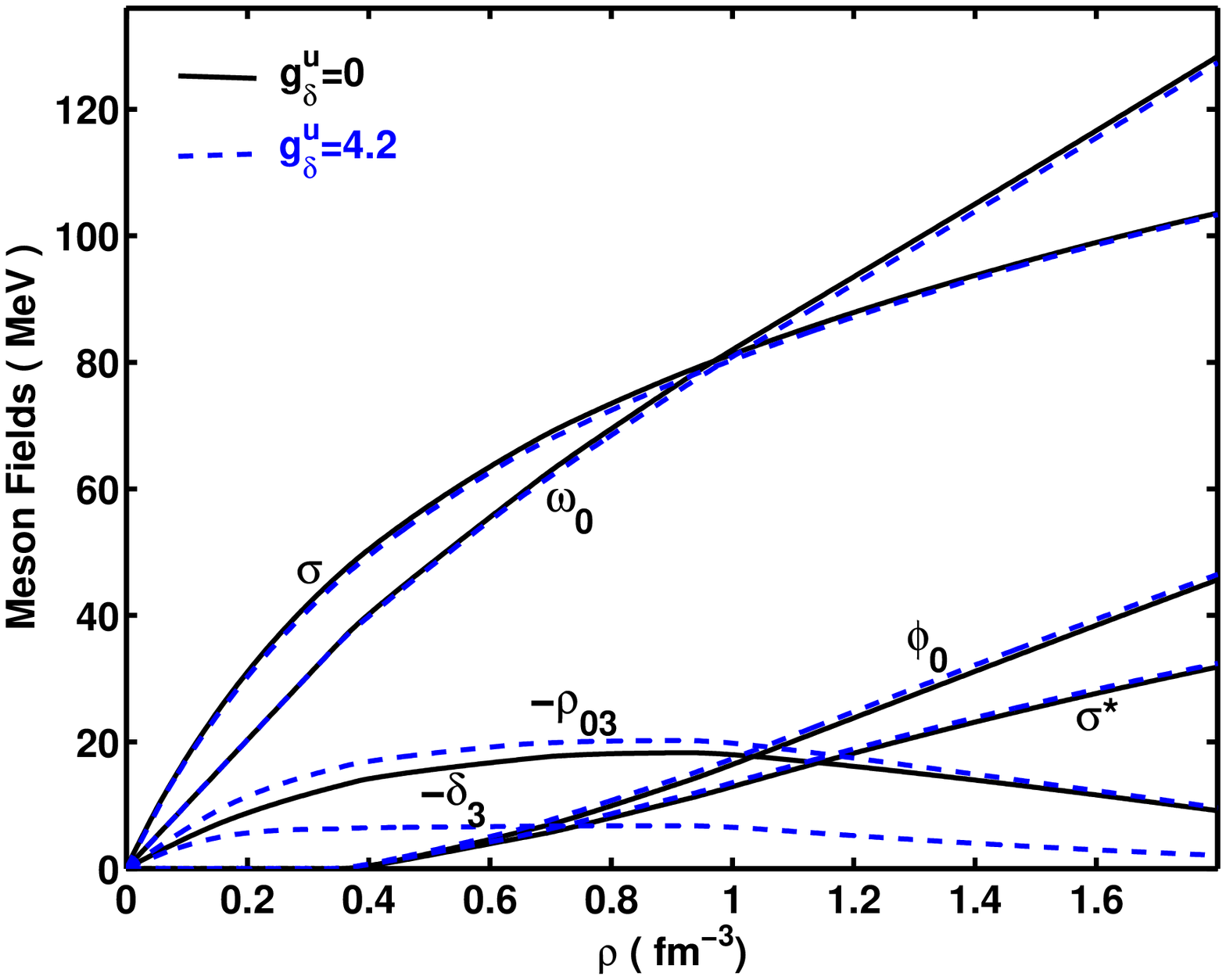}
\includegraphics[width=7.3cm]{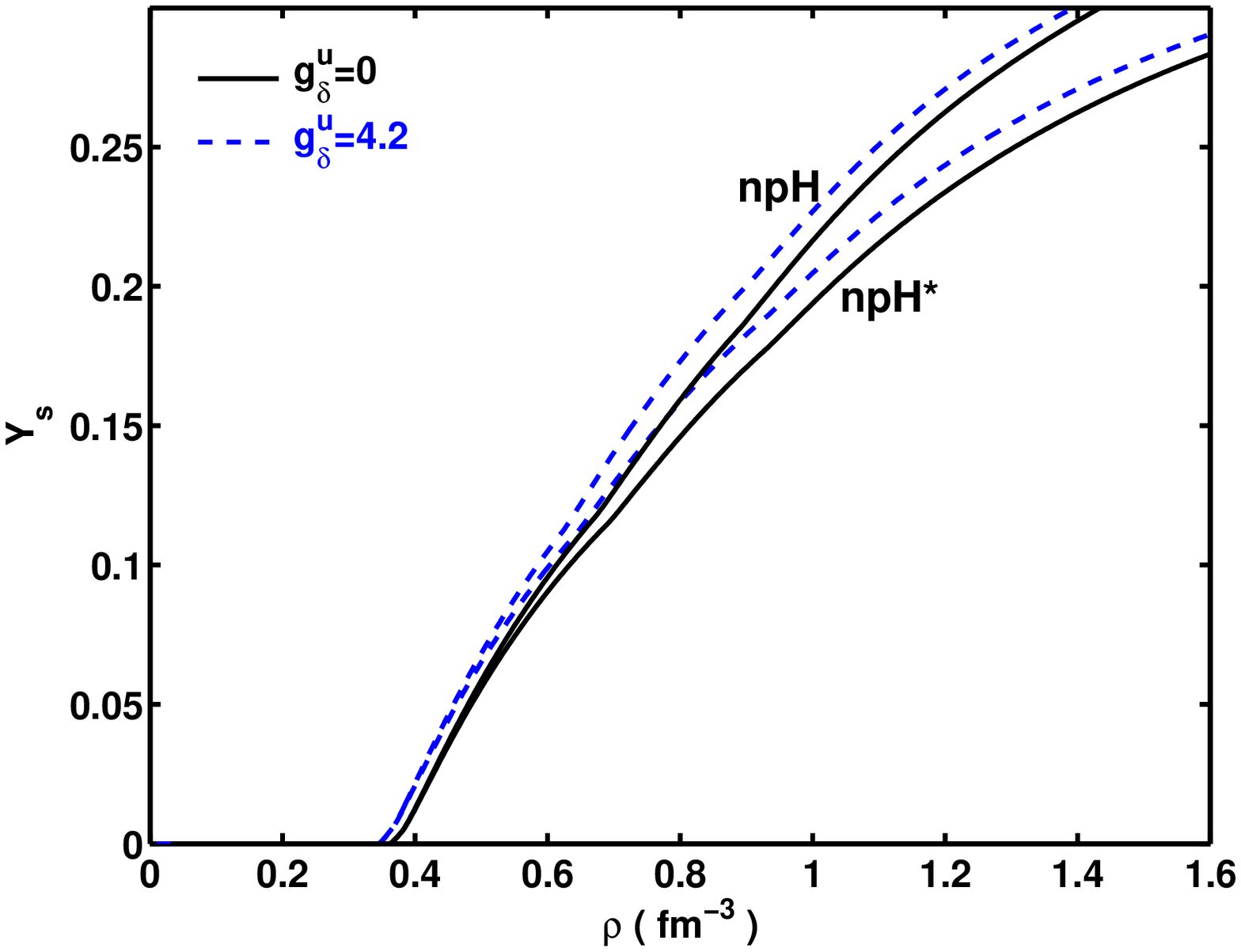}
\caption{(Color online) The left panel are meson fields as functions
of baryon density for $npH^*$. The right panel is the ratio of $s$
quark to total quark in nuclear matter versus baryon density. The
upper two curves are for $npH$ and the lower two are for
$npH^*$.}\label{field}
\end{center}
\end{figure}

The meson fields for $npH^*$ are shown in the left panel of Figure.
\ref{field}. We can see that $\delta$ meson field decreases
$(\sigma, \omega)$ fields while increases $(\sigma^*, \phi)$ fields.
This is because the $\delta$ meson increases the strange number in
nuclear matter, which is shown in the right panel of Figure.
\ref{field}, and $(\sigma^*,\phi)$ couple only to $s$ quark. The
$\delta$ meson increases $\rho_{03}$ meson field, and the effect
becomes smaller when the $\delta$ meson field decreases as baryon
density increases.

\begin{figure}
\begin{center}
\includegraphics[width=7cm]{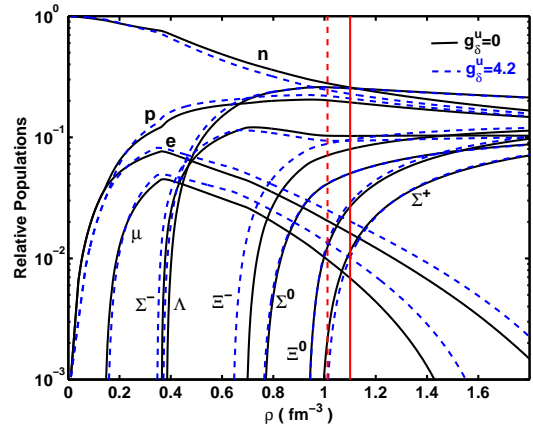}
\caption{The compositions of neutron stars in $npH^*$ as a function
of baryon density and the vertical lines represent the central
baryon densities of neutron stars with maximum
mass.}\label{composition}
\end{center}
\end{figure}

\begin{figure*}
\begin{center}
\includegraphics[width=14cm]{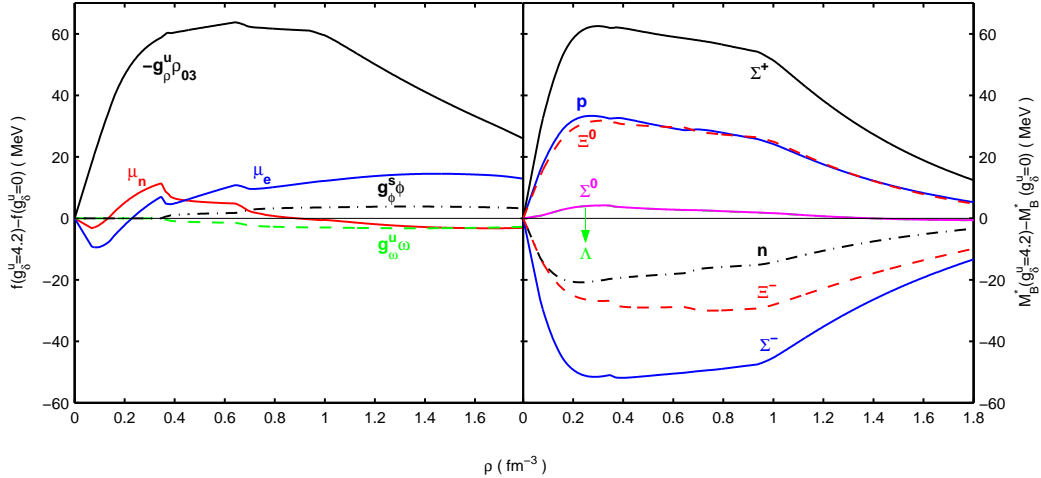}
\caption{(Color online) The left panel is the changes of chemical
potentials of neutron and lepton and the quark-meson exchange
potentials versus the baryon density $\rho$ when the $\delta$ meson
are included. The right panel is the differences of baryon effective
masses. Both are for $npH^*$.}\label{mu}
\end{center}
\end{figure*}

Let's look at the compositions of nuclear matter for $npH^*$ in
Figure. \ref{composition}. $\delta$ meson decreases the neutron
fraction while increases the proton and lepton fractions when
$\rho\gtrsim\rho_0$. From the right panel of Figure. \ref{mu} we see
that $\delta$ meson decreases the effective mass of neutron, which
makes the neutron fraction fall when the density exceed some
critical density which is approximately nuclear matter density
$\rho_0$ as shown in Figure. \ref{composition}. The proton fraction
goes up because the similar reason, and the charge neutrality
condition requires larger lepton fractions.

$\Sigma^{-}, \Xi^{-}$ (negative isospin projection) and $\Lambda,
\Sigma^{0}$ (zero isospin projection) appear earlier when $\delta$
meson are included, but the appearance of $\Sigma^{+}$ (positive
isospin projection) is postponed. From equations (\ref{betaequ}),
(\ref{eB}) and (\ref{BaryonFermi}), we know that the fraction for
baryon $B$ is determined by $(\mu_n-q_B\mu_e)$, $(g_{\omega
B}\omega_0+g_{\phi B}\phi_0+g_{\rho B}I_{3B}\rho_{03})$ and $M_B^*$,
which are all shown in the Figure. \ref{mu}. We see that the changes
of $g_{\omega}^{q} \omega_0$ and $g_{\phi}^{q}\phi$ are proximately
offseted; $ M_B^*$ and $g_{\rho}^u\rho_{03}$ (Compare Figure.
\ref{mu} with Figure. \ref{field}, we can see that the change of
$g_{\rho}^u\rho_{03}$ mainly origins in the large change of
quark-$\rho$ meson coupling constant $g_{\rho}^u$) change obviously.
The changes of $ g_{\rho B}I_{3B}\rho_{03}$ and $M_B^*$ are
isospin-dependent, so whether the hyperon appears earlier is
determined by its isospin projection. The critical density of its
appearance only shifts a little except $\Xi^-$, since the changes of
$ g_{\rho B}I_{3B}\rho_{03}$ and $M_B^*$ are almost the same.

The $\Sigma^-$ population with $\delta$ meson is larger than that
without $\delta$ meson at the beginning, but it becomes smaller than
that without $\delta$ meson because of the appearance of $\Xi^-$.
The reasons are that charge neutrality can be kept more economically
by the larger mass particles with the same charge, and the $\delta$
meson decreases $M_{\Sigma^{-}}^{*}$ more than $M_{\Xi^{-}}^{*}$
(right panel of Figure.\ref{mu}) since the isospin projection of
$\Xi^-$ is $-1/2$ and $\Sigma^-$ is $-1$. $\delta$ meson increases
$\Xi^-$ population obviously larger than other hyperons since it
decreases $\Sigma^-$ population. The appearance of $\Sigma^-$ makes
the lepton fraction begins to fall, which can also be explained by
charge neutrality condition. There is another interesting phenomenon
that the $\Sigma^{+}$ may not appear in neutron stars with $\delta$
meson while its fraction could exceed 1\% for neutron stars at the
maximum masses without $\delta$ meson.

\begin{figure}
\begin{center}
\includegraphics[width=7cm]{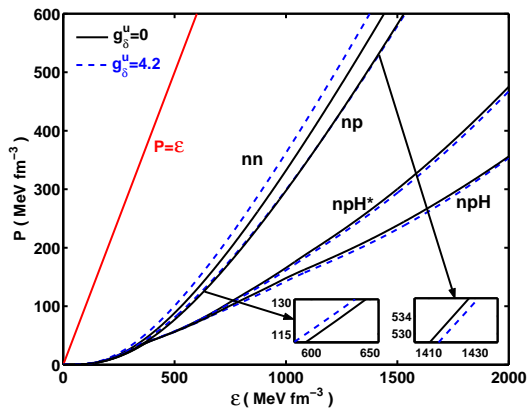}
\caption{(Color online) The equation of state, pressure $P$ versus
energy density $\varepsilon$ for all cases we study in this paper.
The causual limit ($P=\varepsilon$) is also shown.}\label{EOS}
\end{center}
\end{figure}

The EOSes for $nn, np, npH, npH^*$ are plotted in Figure. \ref{EOS}.
The effects of $\delta$ meson can be seen clearly from this figure:
The $\delta$ meson makes the EOS of $nn$ stiffer similar as in QHD
model \cite{EOSH}. For $np$, the $\delta$ meson stiffens the EOS at
low density while softens at high density. The density-dependent
coupling constants are introduced additionally in QHD model
\cite{Liu2} to get the similar results, but the density-dependence
of couplings between scalar mesons and baryons are obtained
self-consistently in our paper. If hyperons are taken into account,
the EOSes with $\delta$ meson suffer a transition to nucleon-hyperon
phase at some density and become softer, this can be seen from the
EOSes of $npH$ and $npH^*$ in Figure. \ref{EOS} clearly. $(\sigma^*,
\phi)$ mesons obviously stiffen the EOSes as in Ref.\cite{Pal}, but
their influences on the $\delta$ meson effect could be neglected.
Since $(\sigma^*,\phi)$ meson fields couple only to $s$ quark and
$\delta$ meson couple only to ($u$, $d$) quarks, $\delta$ meson has
no direct influence on $(\sigma^*,\phi)$ meson fields, which can
also be seen from the lower panel in Figure. \ref{field} as
mentioned above.

We find that no matter whether hyperons are positive, negative or
neutral, their inclusions can make the EOSes with $\delta$ meson
become softer. This result probably reveals that it is the strange
quark makes the EOSes with $\delta$ meson become softer. That is to
say that it is the strange quarks in hyperons results in reversed
direction changes of EOSes if compared with a nucleonic star.
\begin{figure}
\begin{center}
\includegraphics[width=8cm]{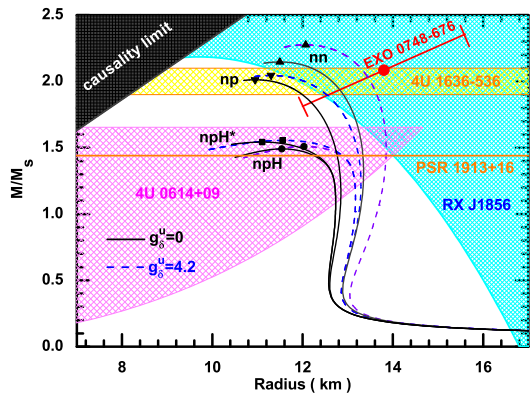}
\caption{(color online) The mass-radius relation for neutron stars
in $nn$, $np$, $npH^*$, $npH$. The filled triangles, inverted
triangles, squares and circles represent the positions of maximum
masses for $nn$, $np$, $npH^*$, $npH$ EOSes, respectively. A
gravitational redshift of 0.35 \cite{redshift} (observed from EXO
0748-676) is shown in this figure. Mass-radius constraints from
thermal radiation of isolated NS RX J1856.5-3754 \cite{RXJ1856}
(cyan hatched area) and from QPOs in the LMXBs 4U 0614+09
\cite{4U0614} (magenta hatched area) and 4U 1636-536 \cite{4U1636}
(orange hatched area) shall be regarded as separate constraints to
the EOSes. The mass of PSR 1913+16 \cite{PSR1913} is also shown in
this figure.}\label{mr}
\end{center}
\end{figure}

The mass-radius relation of neutron stars are shown in Figure.
\ref{mr}. We see that the $\delta$ meson increases the maximum
masses of neutron stars for all cases we studied in this paper. This
is different from QHD model in which the maximum mass decreases for
$np$ with density-dependent couplings \cite{Liu2} and $npH$ when
$\delta$ meson are included \cite{EOSH}. The $\delta$ meson enlarge
the radii of neutron stars about 0.5 km for stars with $M>M_{s}$,
this is an obvious change considering the same EOS at low density
are used for all cases. Another conclusion is that the central
density of neutron star becomes about 0.5 $\rho_0$ smaller when
$\delta$ meson is included. These can be seen from Table \ref{tb4}
clearly. Some observation values are also displayed in Figure.
\ref{mr}. We can see that all cases are compatible with the
observations from PSR 1913+16 \cite{PSR1913} and 4U 0614+09
\cite{4U0614}, but $npH^*$ and $npH$ neutron stars might be ruled
out by neutron star 4U 1636-536 \cite{4U1636} or EXO 0748-676
\cite{redshift}. To show quantitatively the $\delta$ meson effects
on neutron stars properties, the maximum mass $M_{\rm max}$ and the
corresponding radii $R_{M_{\rm max}}$, central baryon density
$\rho_c$, central energy density $\varepsilon_{c}$, central pressure
$P_{c}$ are listed in Table \ref{tb4} for all cases.
\begin{table}
\begin{center}
\caption{The maximum masses of neutron stars and the corresponding
radii $R_{M_{\rm max}}$, central baryon density $\rho_c$, central
energy density $\varepsilon_{c}$, central pressure $P_{c}$ for
different EOSes.}\begin{tabular} {ccccccc}\hline\hline
&$g_\delta^u$&$M_{\rm max}/M_{s}$ &$R_{M_{\rm max}}$  &$\rho_c/\rho_0$ &$\varepsilon_c$ &$P_c$\\
&& &(km) & &(MeV\ fm$^{-3}$) &(MeV\ fm$^{-3}$)\\
\hline
$nn$&4.2 &2.275 &12.07 &5.1 &1093 &418\\
$nn$&0 &2.147 &11.50 &5.6 &1213 &459\\
$np$&4.2 &2.045 &11.30 &6.0 &1274 &446\\
$np$&0 &2.012 &10.95 &6.3 &1352 &493\\
$npH^*$&4.2 &1.556 &11.56 &6.0 &1157 &201\\
$npH^*$&0 &1.543 &11.10 &6.5 &1274 &236\\
$npH$&4.2 &1.509 &12.03 &5.2 &986 &140\\
$npH$&0 &1.491 &11.54 &5.8 &1106 &166\\
\hline \hline
\end{tabular}\label{tb4}
\end{center}
\end{table}
\section{Conclusions}\label{sect5}
We have investigated the $\delta$ meson effects on neutron star
proporties within the modified quark-meson coupling model. We sum up
the conclusions in four aspects:

(1)$\delta\leftrightarrow$strangeness: the $\delta$ meson can make
the pure nucleonic matter EOSes stiffer, while make the hyperon
matter EOSes softer and this could be explained by the appearance of
strange quarks in hyperons.

(2)$\delta\leftrightarrow$leptons: the $\delta$ meson results in
opposite effects on the EOSes of $\beta$-equilibrium nuclear matter
such as $np$, $npH$ and $npH^*$ compared with the EOS of $nn$, which
is similar to the influence of hyperons, but the effect is smaller.
This is because of density-dependence of the couplings between
baryons and scalar mesons.

(3)$\delta\leftrightarrow(\sigma^*, \phi)$: $\delta$ meson has no
direct influence on $s$ quark, so it has little effect on
$(\sigma^*,\phi)$ meson fields. As a result, $(\sigma^*,\phi)$
mesons have no obvious influence to the $\delta$ meson effect on
EOSes, although ($\sigma^*, \phi$) stiffen the EOSes obviously.

(4)$\delta\leftrightarrow$ neutron star properties: the $\delta$
meson can increase the maximum masses of neutron stars, decrease the
corresponding both baryon density and energy density. The radii
become about 0.5 km larger for stars with $M>M_{s}$. It changes
compositions of neutron stars: decrease the neutron fraction and
increase the proton and leptons fractions when $\rho\gtrsim\rho_0$;
make the abundance of $\Xi^{-}, \Xi^{0}$ and $\Sigma^{+}$ larger,
while the abundance of $\Sigma^{-}$ smaller; and increase the
strange number of neutron stars. It can also make the isospin
dependent physical quantities splitting, such as effective baryons
masses.
\section*{Acknowledgments}
One of the authors Chun-Yuan Gao acknowledges the financial support
from the National Natural Science Foundation of China under grants
10305001, 10475002 \& 10435080. The authors are grateful to
Professor Pawel Danielewicz for providing the data for pressure-
density relationship consistent with the experimental flow data
which is indicated by the shaded region in Figure. \ref{gdelta}. We
thank to the useful guidance of Professor Jie Meng and the inspiring
discussions with Chang-Qun Ma and Bao-Yuan Sun.

\end{document}